\documentclass[runningheads]{llncs}

\usepackage[T1]{fontenc}
\usepackage{graphicx}
\usepackage{amsmath}
\usepackage{tabularx}
\usepackage{array}
\newcolumntype{Y}{>{\centering\arraybackslash}m{4cm}} 
\newcolumntype{Z}{>{\centering\arraybackslash}m{3.25cm}} 
\usepackage{booktabs}
\usepackage{lineno}
\usepackage{hyperref}
\usepackage{soul}
\usepackage{xcolor}
\usepackage{ulem}
\usepackage{adjustbox}
\usepackage{makecell}

\sethlcolor{yellow}  

 \makeatletter
\def\@fnsymbol#1{\ensuremath{\ifcase#1\or \dagger\or \ddagger\or
   \mathsection\or \mathparagraph\or \|\or **\or \dagger\dagger
   \or \ddagger\ddagger \else\@ctrerr\fi}}
    \makeatother
\newcommand{\repeatthanks}{\textsuperscript{\thefootnote}}

\begin{document}
\title{Classifier Enhanced Deep Learning Model for Erythroblast Differentiation with Limited Data
}
\titlerunning{Classifier Enhanced DL Model for Erythroblast Diff. with Limited Data}
%
\author{Buddhadev Goswami\thanks{Equal contribution.}\inst{1}\orcidID{0009-0005-5279-066X} \and
Adithya B. Somaraj\repeatthanks\inst{1}\orcidID{0009-0007-7948-777X} \and
Prantar Chakrabarti \inst{2}\orcidID{0000-0003-1026-5427} \and
Ravindra Gudi \inst{3}\orcidID{0000-0002-1655-3887} \and
Nirmal Punjabi\inst{1,4}\orcidID{0000-0002-3953-0936}}

\authorrunning{B. Goswami et al.}


\institute{Koita Centre for Digital Health, Indian Institute of Technology Bombay,\\ Mumbai, India\\
\email{\{buddhadev, adithya, npunjabi\}@iitb.ac.in}  \and
Zoho Corporation, Tenkasi, India\\ 
\email{prantar@gmail.com} \and
Department of Chemical Engineering, Indian Institute of Technology Bombay, Mumbai, India \\
\email{ravigudi@iitb.ac.in} \and
Sensing and Monitoring Foundation, Mumbai, India
}

\maketitle              
\begin{abstract}
Hematological disorders, which involve a variety of malignant conditions and genetic diseases affecting blood formation, present significant diagnostic challenges. One such major challenge in clinical settings is differentiating Erythroblast from  WBCs. Our approach evaluates the efficacy of various machine learning (ML) classifiers—SVM, XG-Boost, KNN, and Random Forest—using the ResNet-50 deep learning model as a backbone in detecting and differentiating erythroblast blood smear images across training splits of different sizes. Our findings indicate that the ResNet50-SVM classifier consistently surpasses other models' overall test accuracy and erythroblast detection accuracy, maintaining high performance even with minimal training data. Even when trained on just 1\% (168 images per class for eight classes) of the complete dataset, ML classifiers such as SVM achieved a test accuracy of 86.75\% and an erythroblast precision of 98.9\%, compared to 82.03\% and 98.6\% of pre-trained ResNet-50 models without any classifiers. When limited data is available, the proposed approach outperforms traditional deep learning models, thereby offering a solution for achieving higher classification accuracy for small and unique datasets, especially in resource-scarce settings.

\end{abstract}

\section{Introduction}

Blood disorders in India present unique challenges due to a range of malignant conditions and genetic diseases affecting blood formation. These include beta thalassemia, hemophilia, iron deficiency anemia, leukemia, lymphoma, etc. Managing these disorders is particularly difficult in a resource-limited setting like India, where these conditions are more prevalent and socio-economically challenging compared to Western countries \cite{abbafhs_classification_2018}. Patients with certain blood disorders may undergo a splenectomy to alleviate symptoms and improve quality of life by preventing excessive destruction of blood cells \cite{https://doi.org/10.1002/ajh.2830380209}. This procedure can increase the presence of nucleated red blood cells (NRBCs) in the blood. NRBCs or erythroblasts are immature red blood cell precursors typically confined to the bone marrow and rarely seen in healthy adults, but they may appear more frequently in post-splenectomy patients \cite{alkafrawi_blood_2022}. Distinguishing NRBCs from lymphocytes in blood smears is challenging due to their similar morphological features. This task is complicated by the variability in lymphocyte appearance and the presence of abnormal cells in hematological disorders. The quality of slide preparation, the microscope's resolution, and the pathologist's expertise are crucial in accurately differentiating these cells \cite{das_hematological_2016}.

\section{Related work}

Das et al. (2016),  identified nucleated red blood cells in 50 blood smear images. Their method integrates multilevel thresholding for cell localization, a unique colour space transformation for enhanced contrast between nucleated cells and RBCs, and special fuzzy c-means clustering for segmentation. A random forest classifier discriminates NRBCs from WBCs with an accuracy of 99.42\%, offering a significant tool for clinicians diagnosing various anemic conditions efficiently \cite{das_hematological_2016}. \\ Fang et al. (2022) in their study present a novel, label-free technique for identifying rare NRBC using deep learning and single-cell Raman spectroscopy. By combining Faster RCNN and YOLOv3 for morphological detection and Raman for verification, it offers rapid, efficient NRBC screening without pre-processing \cite{fang_fast_2022}. Alkafrawi et al. (2023) developed an AlexNet-based Convolutional Neural Network model that classifies and counts blood cells in microscopic images with 95.08\% accuracy using a dataset of 17,092 blood smear samples. This model showcases the effectiveness of deep learning in medical diagnostics. Additionally, they created a user-friendly GUI, `Blood Cell Classifier v1.0,' to help hematologists classify blood cells efficiently, illustrating how machine learning can automate traditional manual counting methods \cite{alkafrawi_blood_2022}.  Rao et al. (2023) proposed  EfficientNet - XGBoost framework as a novel method for segmenting and classifying white blood cells (WBCs) from 367 blood smear images. This method uses SegNet for segmentation, EfficientNet for feature extraction, and XGBoost for classification, achieving a higher rank-1 accuracy of 99.02\% compared to existing techniques.\cite{sivarao_efficientnet_2023}
Chola et al. (2022) proposed that BCNet is a deep learning model aiming to improve accuracy of blood cell classification by identifying multiclass blood cells rapidly and automatically suing 17,029 mages. The model uses ResNet-18 as the backbone model, achieving 96.78\% accuracy \cite{chola_bcnet_2022}.
 Nozaka et al. (2024) used ResNet models identifying immature granulocytes(IG) and erythrocytes while screening peripheral blood smears using 6727 images. Deep learning techniques and  The findings demonstrated a precision level of 97\% for healthy cases and 88\% for cases with immature granulocytes. The  model for IG recognition, based on CNN, achieved an accuracy rate of 97\% for healthy cases and 88\% for IG cases\cite{nozaka} 

 While all the above papers worked on blood cell classification, none focused on data-efficient learning for detection and differentiation of NRBC v/s other WBBCs. Our study addresses this problem and compares various ML classifiers across various splits. 


\section{Dataset}

\begin{figure} [!h]
\includegraphics[width=\textwidth]{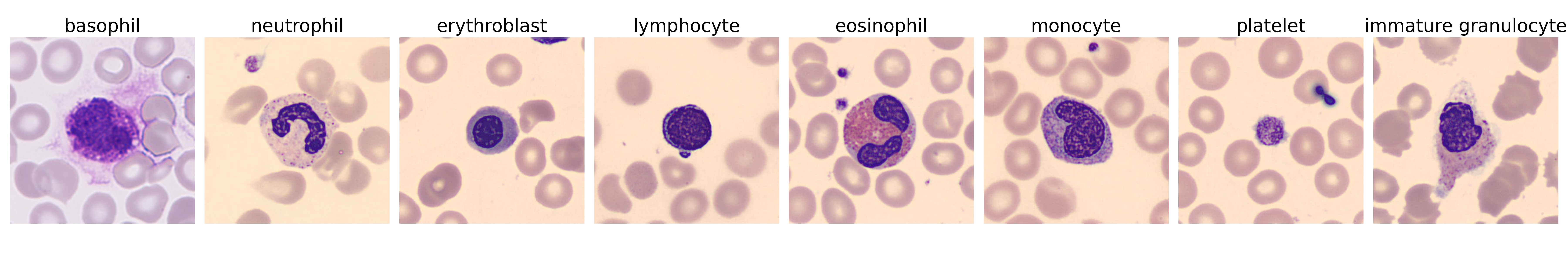}
\caption{Sample images from each class of the dataset.} \label{Fig1}
\end{figure}

For this study, the dataset was sourced from the Mendeley repository called `A dataset for microscopic peripheral blood cell (PBC) images for development of automatic recognition systems' \cite{acevedo_dataset_2020}. This dataset comprises excellent digital images of typical peripheral blood cells. The images were obtained using the CellaVision DM96 analyzer at the Hospital Clinic of Barcelona after conducting cell preparation and staining procedures with the Sysmex SP1000i and May Grünwald-Giemsa stain, respectively. The dataset comprises 17,092 JPEG images with dimensions of 360x363 pixels. It is organized into eight distinct categories of blood cells: neutrophils, eosinophils, basophils, lymphocytes, monocytes, immature granulocytes (including metamyelocytes, myelocytes, and promyelocytes), erythroblasts, and platelets as shown in Fig. \ref{Fig1}. 

This study specifically required a dataset containing erythroblasts and lymphocytes for effective differentiation. The Mendeley dataset uniquely meets this requirement, as other accessible datasets like LISC\cite{rezatofighi_automatic_2011} or ALL-IDB\cite{6115881} do not include the erythroblast class essential for our analysis.

\section{Experimental Methodology}

 All the experiments were conducted on a single Nvidia GeForce RTX-3050 GPU device with 8 GB of RAM. We have used a batch size of 64 images, leveraging CUDA libraries for optimized performance.

\subsection{Backbone Selection }

A range of pre-trained convolutional neural network architectures were selected for our classification tasks, including ResNet-50 \cite{he_deep_2015}, VGG19 \cite{simonyan_very_2015}, ResNet-18, VGG16, and InceptionV3 \cite{szegedy_going_2014}. The dataset was divided into 70\% for training, 15\% for validation, and 15\% for testing. All models underwent a 15-epoch training regimen, except InceptionV3, which required 30 epochs due to its more complex structure. According to the initial evaluation results presented in Table \ref{tab:model_performance}, ResNet-50 demonstrated the highest testing accuracy at 98.72\%, leading us to select it as our primary feature extraction backbone. Unlike the VGG models, ResNet-50 features residual connections that mitigate the vanishing gradient problem, facilitating more efficient training and enabling the construction of deeper, more effective networks\cite{he_deep_2015}. As seen in Table \ref{tab:model_performance}, ResNet-50 provides highest accuracy with second lowest training time. 
ResNet-50's optimal balance of depth and computational efficiency allows it to handle complex features more effectively than ResNet-18 while avoiding the greater computational demands of higher ResNet models. Additionally, ResNet-50's extensive availability of pretrained models makes it highly suitable for efficient transfer learning and deployment. In contrast to newer models like 
EfficientNet \cite{sivarao_efficientnet_2023}, DenseNet, and Vision Transformers (ViT), ResNet-50 remains advantageous for several reasons. EfficientNet, while highly efficient due to compound scaling, presents complexity in understanding and implementation. DenseNet's dense connectivity enhances feature reuse and gradient flow but incurs higher memory and computational costs\cite{huang2018denselyconnectedconvolutionalnetworks}. ViT excels in capturing long-range dependencies and performs well on large-scale datasets but demands extensive data and computational resources\cite{dosovitskiy2021imageworth16x16words}. Despite the advancements in these newer models, ResNet-50 remains a balanced choice, offering robustness, efficiency, and accessibility for diverse real-world applications.
\begin{table}[htbp]
\centering
\caption{Model performance comparison for backbone selection with 70:15:15 splits}\label{tab:model_performance}
\renewcommand{\arraystretch}{1.5} 
\begin{adjustbox}{max width=\textwidth}
\begin{tabular}{|l|c|c|c|c|c|c|}
\hline
\textbf{Model} & \makecell{\textbf{Test} \\ \textbf{Accuracy}} & \makecell{\textbf{Precision}} & \makecell{\textbf{Recall}} & \makecell{\textbf{F1} \\ \textbf{Score}} & \makecell{\textbf{Trainable} \\ \textbf{Parameters}} & \makecell{\textbf{Total Training} \\ \textbf{Time}} \\
\hline
ResNet-50   & 98.72\% & 0.9892 & 0.9874 & 0.9882 & 25 million & 23m 24s \\
\hline
VGG-16      & 98.67\% & 0.9826 & 0.9834 & 0.9832 & 138 million & 47m 33s \\
\hline
ResNet-18   & 98.57\% & 0.9862 & 0.9868 & 0.9892 & 11 million & 14m 15s \\
\hline
VGG-19      & 98.44\% & 0.9852 & 0.9864 & 0.9854 & 143 million & 53m 33s \\
\hline
InceptionV3 & 92.38\% & 0.9179 & 0.9171 & 0.9186 & 23 million & 37m 56s \\
\hline
\end{tabular}
\end{adjustbox}
\end{table}


\subsection{Training on ResNet-50 architecture }

ResNet-50 architecture, tailored for image classification, was implemented using the PyTorch library for building the model, and torch-vision was used for data pre-processing. The dataset was organized into distinct classes and divided into training, validation, and testing.
The testing set was fixed with 4000 images, and the other sets were varied as per the data split.

In this setup, the ResNet-50 model, initially trained on the ImageNet\cite{imagenet} dataset, was fine-tuned by adjusting its final layers for blood cell classification.  The cross-entropy loss function directs the training process is given in equation \ref{eq2}
\begin{equation}
\label{eq2}
    L(y, \hat{y}) = -\sum y \log(\hat{y})
\end{equation}
which is the most effective choice for multi-class problems. This loss function evaluates the model's output by comparing it to the actual data labels. The Adam optimizer is utilized for optimization, as it is known for its efficiency in adaptive updating network weights. \par

Before we began the training process, the optimal learning rate was determined by gradually increasing the learning rate over a predefined range. At each iteration, the training loss was recorded, and we chose the optimal learning rate. This is the point at the middle of the steepest downward curve, just before divergence. Following this, we implemented discriminative fine-tuning\cite{howard2018universallanguagemodelfinetuning}and assigned higher learning rates to later layers in the model, while earlier layers have progressively lower learning rates. Using the optimal learning rate identified earlier as the maximum for the final layer, we applied a one-cycle learning rate scheduler. This scheduler starts with a small initial learning rate, increases it to the maximum, and then decreases it to a final value lower than the initial rate. Then, we trained our model using k-fold cross-validation with a 5-fold configuration over 15 epochs. Each fold of the validation data is used once as a test set, while the entire training dataset is used to train the model in each epoch. The validation indices are shuffled to randomize the data selection, ensuring unbiased validation subsets. The model is then evaluated on each validation subset, with performance metrics such as loss and accuracy (both top-1 and top-5) calculated and stored. After evaluating all folds, the mean and standard deviation of these metrics are computed to assess overall model performance and consistency across different subsets. If the average validation loss of a fold is lower than previously recorded, the model's state is saved. This process helps in selecting the model configuration that generalizes best on unseen data.

\subsection{Classifiers  }
A hybrid methodology that combines deep learning and traditional machine learning techniques to tackle image classification tasks was adopted. The computational capabilities of PyTorch library\cite{paszke2019pytorch}, supplemented by algorithms from Scikit-learn\cite{pedregosa2018scikitlearn}  enhanced efficiency.

A pre-trained ResNet-50 model was utilized and adapted for our classification task. The pre-trained network acts as a feature extractor where the initial layers capture generic features (edges, textures) while deeper layers identify more complex patterns relevant to the specific classes in our dataset. The last fully connected layer, reformulated for our needs, transforms these features into class probabilities using the softmax function shown in equation \ref{eq3}

\begin{equation}
 \label{eq3}
 {Softmax}(z_i) = \frac{e^{z_i}}{\sum_{j=1}^{K} e^{z_j}}
\end{equation}

where $z_i$ are the logits (i.e., unnormalized log probabilities) produced by the last network layer for each class, and $K$ is the total number of classes. Throughout the training and evaluation phases, we meticulously monitored various performance metrics, such as accuracy, precision, recall, and F1-score, to fine-tune and evaluate our models. The dataset was systematically sorted into training, validation, and testing directories. Preproccesing is done, which included uniform image transformations such as resizing to $224 \times 224$ pixels, center cropping, converting images into tensor format, and normalizing based on the mean ($\mu$) and standard deviation ($\sigma$) values derived from the ImageNet dataset\cite{imagenet} shown in equation \ref{eq4}.

\begin{equation}
 \label{eq4}
 {Normalized} = \frac{\text{Image} - \mu}{\sigma}
\end{equation}

where $\mu = [0.485, 0.456, 0.406]$ and $\sigma = [0.229, 0.224, 0.225]$ (values obtained from ImageNet). These preprocessing steps were essential for preparing the data for optimal processing through neural network architectures.
Additional preprocessing or image augmentation was not performed due to the nature of the data, as it is a well-curated dataset.

\paragraph{Our modeling strategy }involved dual approaches: fine-tuning traditional machine learning models (KNN\cite{KNN}, SVM\cite{SVM}, RandomForest\cite{RandomForest}, XGBoost\cite{XGboost}) and adapting a deep learning model. We employed grid search\cite{shekar} to optimize the hyperparameters of the traditional models, shown in equation \ref{eq5}.

\begin{equation}
 \label{eq5}
 {Grid Search} = \arg\max_{\theta} \left(\sum_{i=1}^{n} \text{Accuracy}(\theta_i)\right)
\end{equation}

\begin{figure}[ht]
\includegraphics[width=\textwidth]{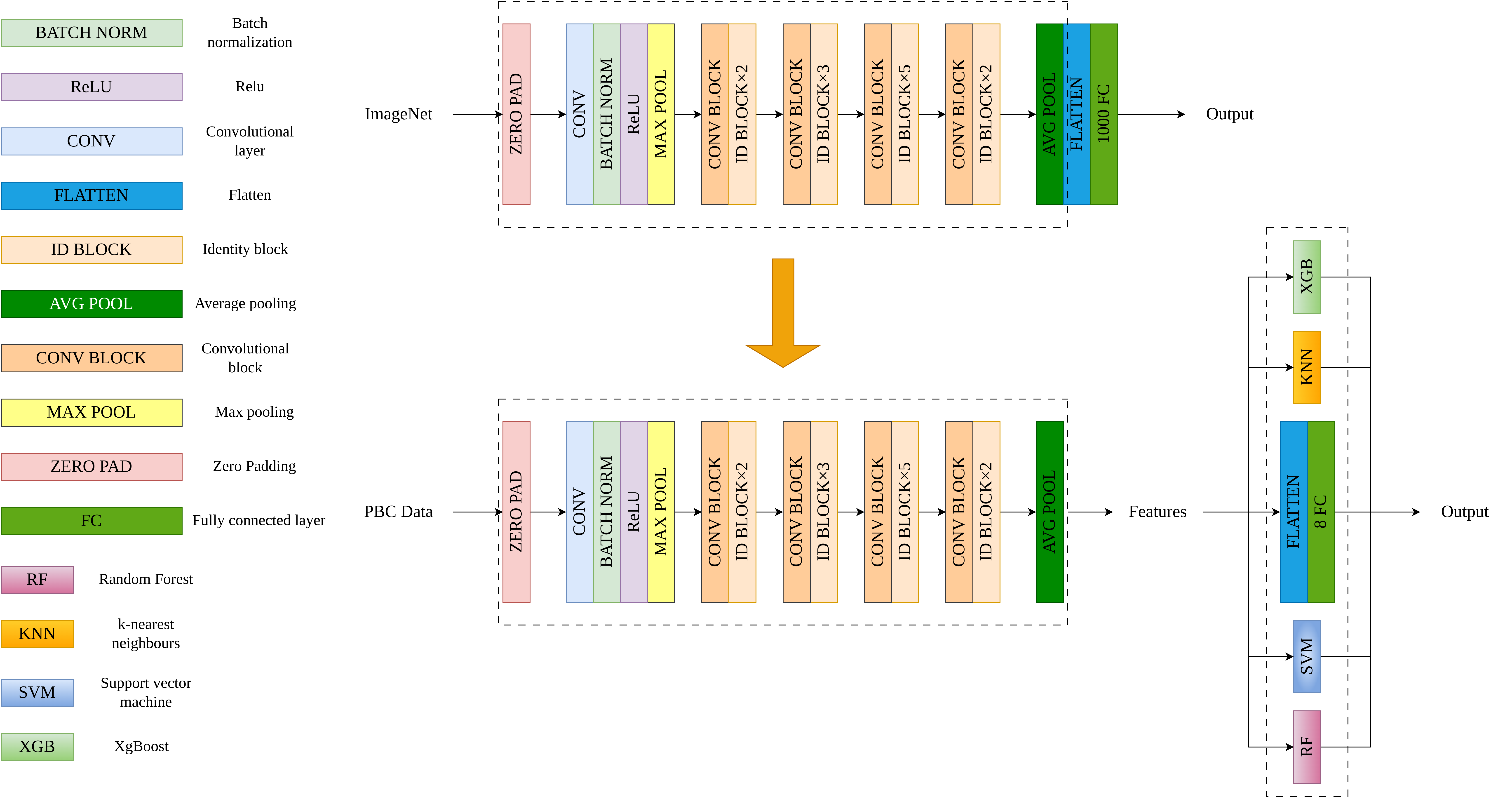}
\caption{Our proposed classifier enhanced ResNet-50 model architecture for the study.} \label{fig2}
\end{figure}

Where $\theta$ represents the set of parameters over which the search is conducted, and $n$ is the number of parameter combinations tested. This exhaustive parameter optimization ensured that our models were highly tailored to maximize accuracy on our specific dataset. Figure \ref{fig2} shows our proposed Architecture for the study.\\

The two-step process of feature extraction followed by classifier implementation, though time-consuming, is crucial for maximizing accuracy with limited data. This method and the application of grid search for optimizing classifier parameters extend the training time. However, these steps are essential for achieving the high precision necessary in clinical applications, especially under conditions of data scarcity. Despite the potential increase in training time due to these methods, the GPU requirements do not substantially change. The computational resources required remain consistent with tasks, making our approach feasible within typical clinical research settings.

\subsection{Training with classifiers}

To assess our classifiers' performance under limited training data conditions, we devised a series of experiments with varied training set sizes, ensuring that each class was represented equally across all partitions.  Table \ref{tab:distribution} provides the information above the dataset distribution.

\begin{table}[!h]
\centering
\caption{Distribution of Training, Validation, and Testing Images}\label{tab:distribution}
\begin{tabularx}{\textwidth}{|X|X|X|X|}
\hline
\textbf{Training Percentage} & \textbf{Training Images} & \textbf{Validation Images} & \textbf{Testing Images} \\ \hline
1\%                          & 168                         & 12,924                         & 4,000                    \\ \hline
2.5\%                        & 424                         & 12,668                         & 4,000                    \\ \hline
5\%                          & 848                         & 12,244                         & 4,000                    \\ \hline
7.5\%                        & 1,273                       & 11,819                         & 4,000                    \\ \hline
10\%                         & 1,709                       & 11,383                         & 4,000                    \\ \hline
20\%                         & 3,418                       &  9,674                         & 4,000                    \\ \hline
30\%                         & 5,128                       &  7,964                         & 4,000                     \\ \hline
\end{tabularx}
\end{table}




\

\begin{table}[htbp]
\centering
\caption{Detailed Test Performance Metrics Across Various Classifiers with a ResNet-50 Backbone}\label{tab:metrics}
\begin{tabularx}{\textwidth}{@{}lYZZ@{}}  
\toprule
Data Split & Classifier & Test Acc (\%) & Erythroblast       (Prec./Rec./ F1) \\
\midrule
\multicolumn{4}{c}{1\% Data Split} \\
\midrule
1\%        & ResNet50                 & 82.03 & 0.986 / 0.857 / 0.917 \\
1\%        & ResNet50-XGBoost         & 84.05 & 0.986 / 0.857 / 0.917 \\
1\%        & ResNet50-KNN             & 85.88 & 0.984 / 0.819 / 0.894 \\
1\%        & ResNet50-SVM             & 86.75 & 0.989 / 0.812 / 0.892 \\
1\%        & ResNet50-RandomForest    & 84.68 & 0.993 / 0.763 / 0.863 \\
\midrule
\multicolumn{4}{c}{2.5\% Data Split} \\
\midrule
2.5\%      & ResNet50                 & 86.25 & 0.990 / 0.794 / 0.881 \\
2.5\%      & ResNet50-XGBoost         & 86.82 & 0.988 / 0.844 / 0.910 \\
2.5\%      & ResNet50-KNN             & 87.55 & 0.976 / 0.882 / 0.926 \\
2.5\%      & ResNet50-SVM             & 88.02 & 0.976 / 0.900 / 0.937 \\
2.5\%      & ResNet50-RandomForest    & 87.20 & 0.986 / 0.866 / 0.922 \\
\midrule
\multicolumn{4}{c}{5\% Data Split} \\
\midrule
5\%        & ResNet50                 & 92.99 & 0.979 / 0.928 / 0.953 \\
5\%        & ResNet50-XGBoost         & 92.77 & 0.969 / 0.928 / 0.948 \\
5\%        & ResNet50-KNN             & 93.05 & 0.949 / 0.926 / 0.937 \\
5\%        & ResNet50-SVM             & 93.00 & 0.953 / 0.926 / 0.939 \\
5\%        & ResNet50-RandomForest    & 91.90 & 0.981 / 0.924 / 0.952 \\
\midrule
\multicolumn{4}{c}{7.5\% Data Split} \\
\midrule
7.5\%      & ResNet50                 & 96.29 & 0.984 / 0.970 / 0.977 \\
7.5\%      & ResNet50-XGBoost         & 95.40 & 0.878 / 0.980 / 0.926 \\
7.5\%      & ResNet50-KNN             & 95.93 & 0.990 / 0.958 / 0.974 \\
7.5\%      & ResNet50-SVM             & 96.07 & 0.988 / 0.960 / 0.974 \\
7.5\%      & ResNet50-RandomForest    & 95.23 & 0.986 / 0.952 / 0.969 \\
\midrule
\multicolumn{4}{c}{10\% Data Split} \\
\midrule
10\%       & ResNet50                 & 96.14 & 0.984 / 0.972 / 0.978 \\
10\%       & ResNet50-XGBoost         & 95.87 & 0.984 / 0.958 / 0.971 \\
10\%       & ResNet50-KNN             & 96.25 & 0.988 / 0.958 / 0.973 \\
10\%       & ResNet50-SVM             & 96.00 & 0.990 / 0.950 / 0.969 \\
10\%       & ResNet50-RandomForest    & 95.70 & 0.972 / 0.964 / 0.968 \\
\midrule
\multicolumn{4}{c}{20\% Data Split} \\
\midrule
20\%       & ResNet50                 & 97.66 & 0.986 / 0.974 / 0.980 \\
20\%       & ResNet50-XGBoost         & 96.25 & 0.980 / 0.958 / 0.969 \\
20\%       & ResNet50-KNN             & 97.48 & 0.967 / 0.988 / 0.977 \\
20\%       & ResNet50-SVM             & 97.30 & 0.956 / 0.990 / 0.973 \\
20\%       & ResNet50-RandomForest    & 96.55 & 0.986 / 0.974 / 0.980 \\
\midrule
\multicolumn{4}{c}{30\% Data Split} \\
\midrule
30\%       & ResNet50                 & 98.36 & 0.982 / 0.996 / 0.989 \\
30\%       & ResNet50-XGBoost         & 97.88 & 0.990 / 0.996 / 0.993 \\
30\%       & ResNet50-KNN             & 98.45 & 0.980 / 0.996 / 0.988 \\
30\%       & ResNet50-SVM             & 98.42 & 0.982 / 0.996 / 0.989 \\
30\%       & ResNet50-RandomForest    & 98.00 & 0.976 / 0.994 / 0.985 \\
\bottomrule
\end{tabularx}
\end{table}

\section{Results}

The performance of ResNet-50, both with and without classifiers like SVM, XG-Boost, KNN, and Random Forest, is compared in terms of test accuracy and erythroblast identification across different training dataset sizes, as shown in Table \ref{tab:metrics}. Remarkably, ResNet-50-KNN outshone ResNet-50 at very low data splits in test accuracy, securing 85.88\% with just 1\% of data. This performance advantage was maintained as dataset sizes increased. In the critical task of erythroblast identification, which is key for precise early detection of red blood cells, ResNet-50-KNN achieved a remarkable precision of 0.989 and a recall of 0.819 at the same minimal data size, aligning closely with ResNet-50's metrics.

As the data size expanded, all classifiers saw enhancements in erythroblast accuracy, with ResNet-50-SVM and Random Forest demonstrating notable proficiency; both achieved F1-scores exceeding 0.973 and 0.980, respectively, at a 20\% data size. ResNet-50-XGBoost exhibited sturdy performance, attaining a test accuracy of 96.25\% at a 20\% data size. ResNet-50-KNN also displayed substantial gains as data volume grew, offering accuracy comparable to other models and erythroblast metrics.

\begin{figure}
\includegraphics[width=\textwidth]{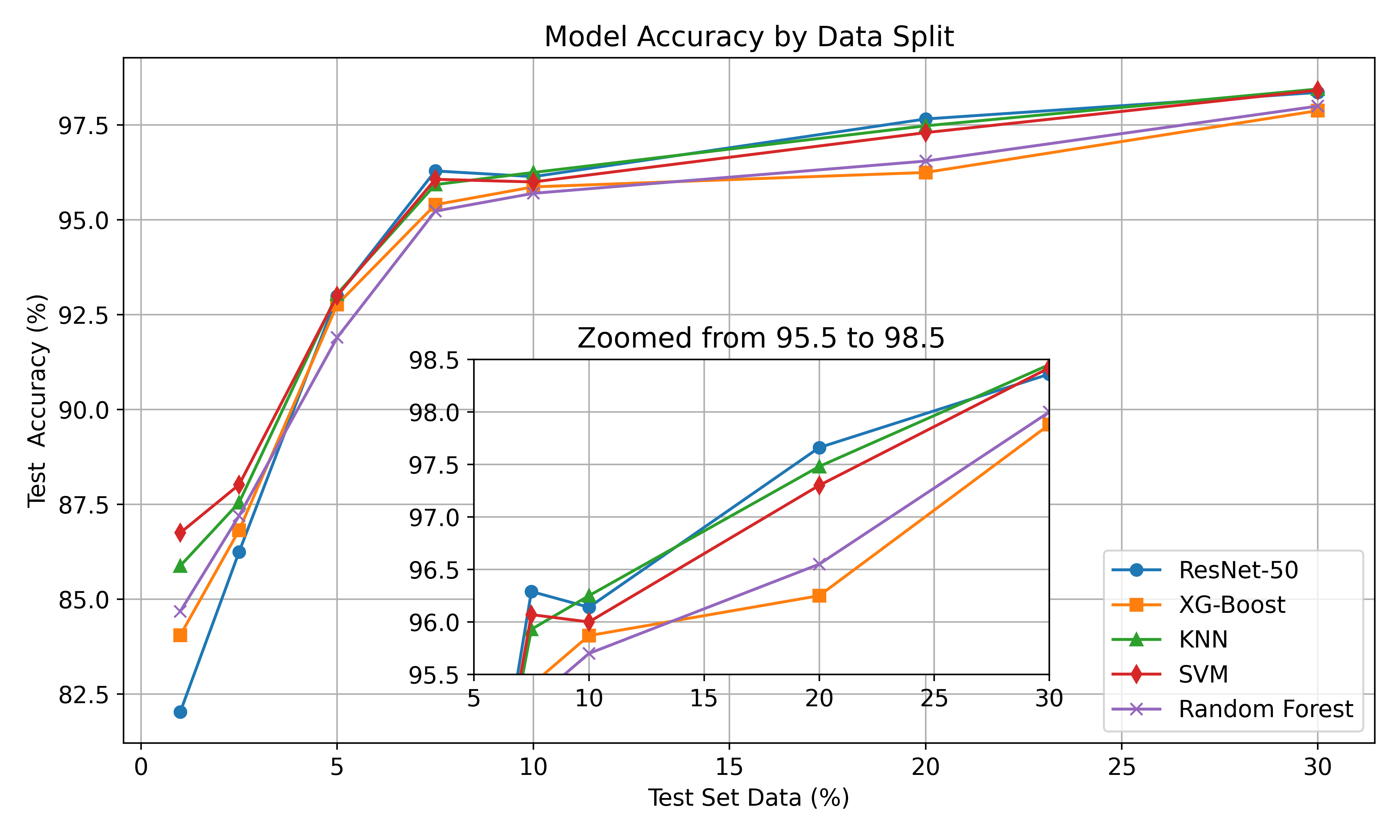}
\caption{Test accuracy V/S percentage of the test data for the various models. Inset shows the differences between the model for higher test data} \label{fig3}
\end{figure}

ResNet-50-SVM outperforms the ResNet-50 model in low data and shows consistent performance, as shown in Fig. \ref{fig3}.

\section{Discussion }
Collaboration with clinical hematologists and pathologists has improved our understanding of the capabilities and limitations of machine learning (ML) algorithms for blood cell classification. These models, notably the one with Support Vector Machine (SVM) as the classifier, achieved a test accuracy of 86.75\% using only 1\% of the available data. This success indicates ML's potential to enhance hematological diagnostics, which is critical for early detection and management of diseases like leukemia and anemia.

Utilizing SVM or alternative machine learning classifiers like KNN with features derived from ResNet-50 results produces superior outcomes in low training data than a standard pre-trained ResNet-50 model. ResNet-50 offers superior features that can efficiently be utilized by SVMs, enabling thorough customization of the dataset's unique characteristics through fine-tuning of hyperparameters and kernel selection. Support Vector Machines (SVMs) adaptability allows them to match the data more effectively than the fixed, fully connected layers commonly present in pre-trained models. Moreover, utilizing ResNet-50 mainly for extracting features minimizes the likelihood of overfitting, particularly in datasets with limited size. SVMs exhibit improved generalization capabilities on unfamiliar data due to their controlled training dynamics. Integrating ResNet-50's advanced deep learning capabilities with the precise adaptability of machine learning classifiers such as SVMs significantly enhances performance.

However, in conventional clinical settings, these models face challenges due to biological variability and limitations in training datasets. Performance issues often arise with images of overlapping cells, a common scenario in clinical samples. This results in frequent misclassifications and underscores the need for advanced image segmentation algorithms to isolate individual cells effectively. Additionally, the models were trained on highly cropped and zoomed images, further limiting their application to typical clinical images. Variability in staining methods and slide quality also affect model performance, as these factors can alter the appearance of cells on slides.

\begin{figure}[!h]
    \centering
    \includegraphics[width=\textwidth,height=5cm,keepaspectratio]{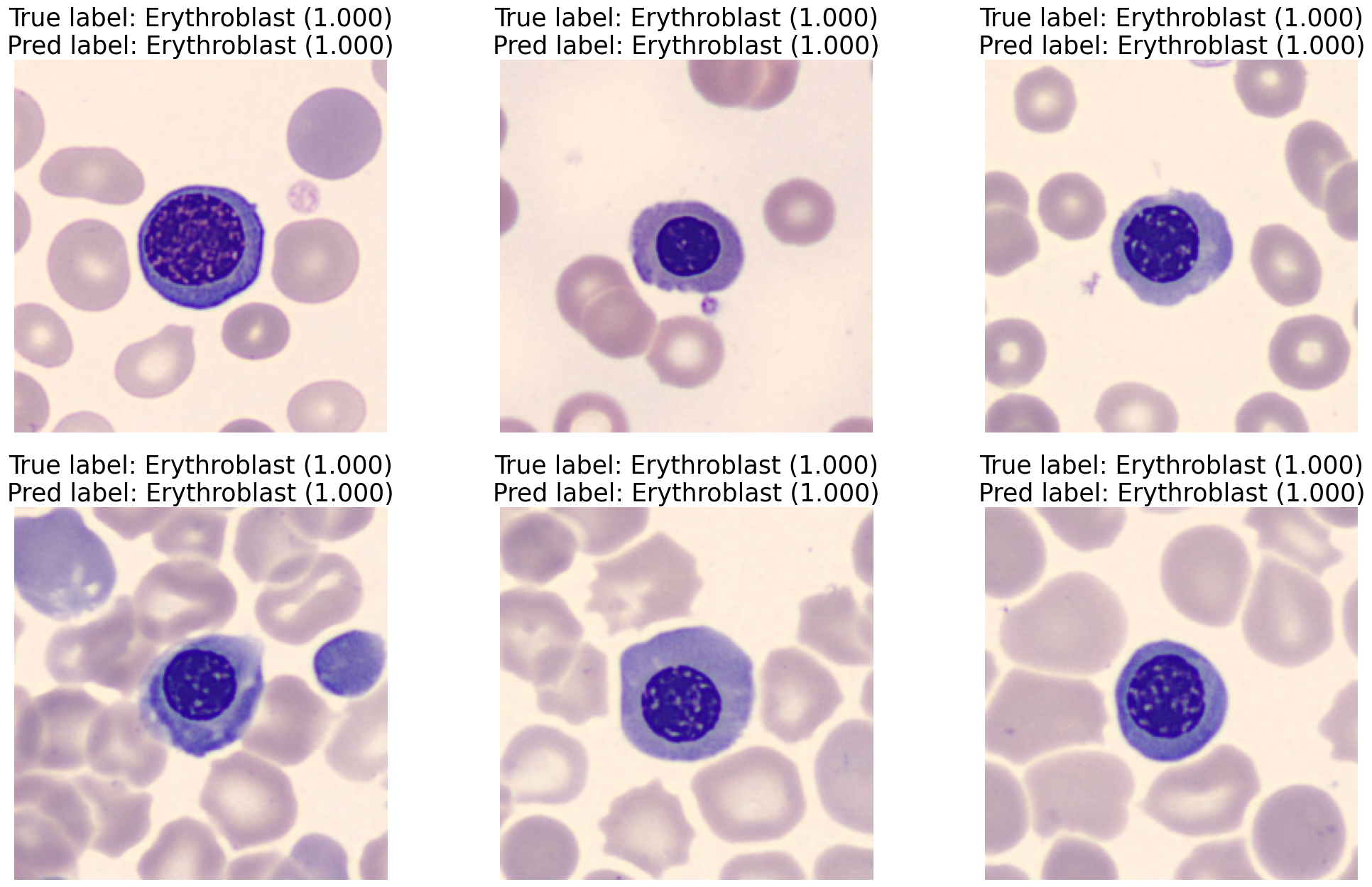}
    \caption{Correctly predicted labels of erythroblast}
    \label{correct}
\end{figure}

\begin{figure}[!h]
    \centering
    \includegraphics[width=\textwidth,height=5cm,keepaspectratio]{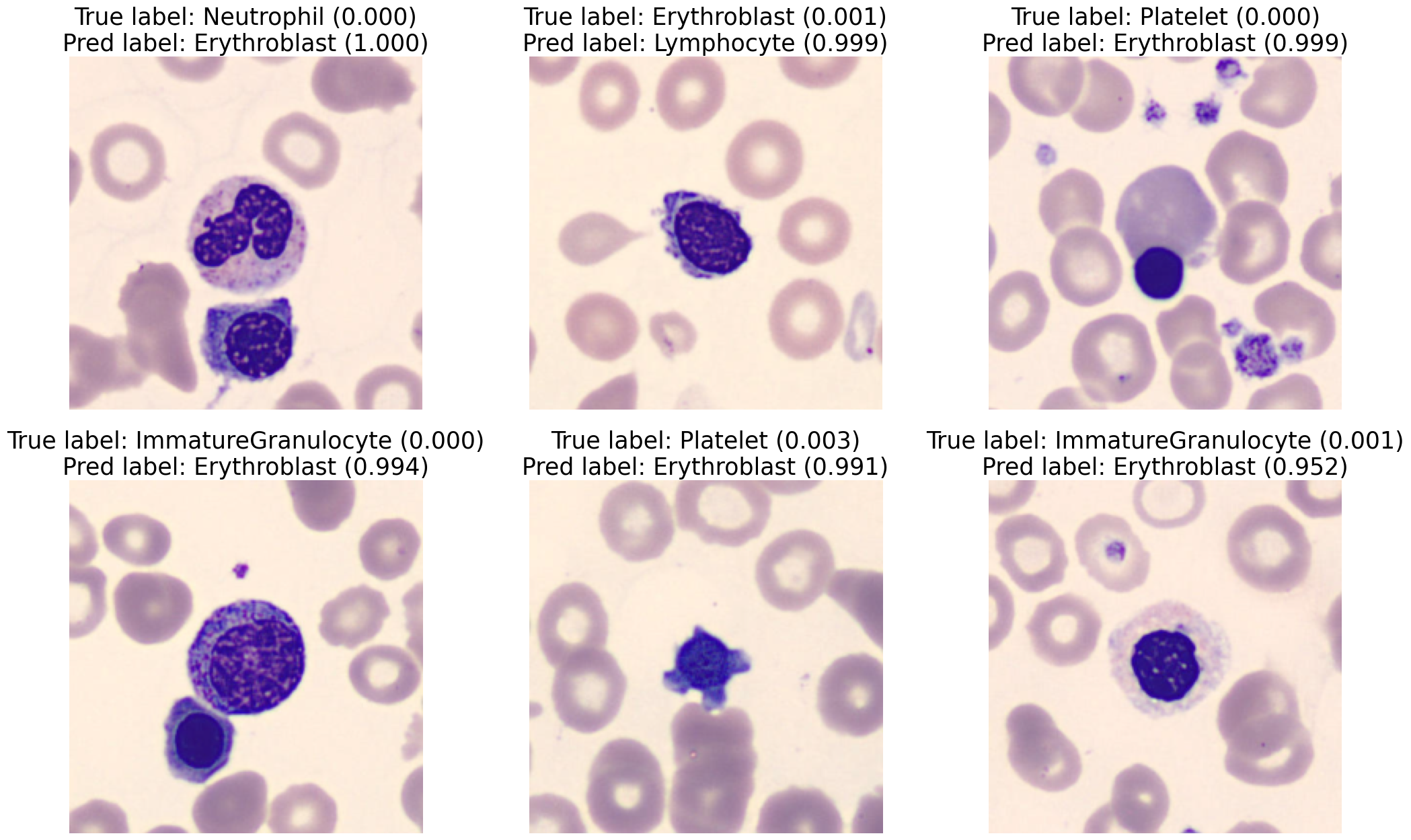}
    \caption{Incorrect predicted labels of erythroblast}
    \label{incorrect}
\end{figure}

The model demonstrates high accuracy in detecting erythroblasts in cases where there is a solitary cell with a clearly defined ratio between the nucleus and cytoplasm in the image, as shown in Figure \ref{correct}. However, it faces difficulties in accurately categorizing erythroblasts in situations where there are multiple cells in a single image, the ratio of nucleus to cytoplasm is low, and small cells like platelets are next to red blood cells (giving the appearance of a single cell), or other cells show visible cytoplasm as shown in Figure \ref{incorrect}.

For the following work, there will be two models in conjunction. The first will segment individual cells from complex histopathology images. The proposed architecture in this study will help classify individual cells to further improve the performance. There is a need to expand the dataset diversity, and training models on comprehensive image data from patients with hematological disorders is essential. Developing a user-friendly interface that normalizes erythroblast counts relative to other white blood cells (WBCs) through ratio calculations or percentage conversions will facilitate easier comparisons across samples. These advancements will enhance model accuracy and robustness, reduce pathologists' workloads, and improve diagnostic processes, bridging the gap between theoretical precision and practical usability in clinical settings.


\section{Conclusion}
This study highlights the significant impact of machine learning models in hematological diagnostics. It showcases the ability of  algorithms, such as SVM and ResNet-50, to accurately classify blood cells with limited images available for training, emphasizing their transformative potential. Although controlled testing environments have achieved high accuracy, real-world clinical applications face significant challenges due to biological variability and limitations in dataset diversity. The highlighted concerns encompass the potential for misclassification due to the proximity of cells in blood smears, as well as the influence of inconsistent staining quality on the model's performance. To tackle these challenges, it is necessary to improve the representativeness of the dataset and refine image processing techniques to ensure that the model performs well in different clinical settings.

Furthermore, the need to train these models with a limited amount of data brings attention to wider concerns regarding fairness in global health—specifically, the challenges imposed by limited resources in regions such as India. In such contexts, efficient models necessitating less data are crucial, as they provide scalable solutions that rapidly adjust to various medical environments without requiring extensive computational resources.

\subsubsection{Acknowledgements}
We thank Dr Pronati Gupta, Consultant Hematopathologist, Chittaranjan National Cancer Institute, Kolkata, for her feedback.

\subsubsection{Code and Data Availability}
The code and data used in this study are available in this \href{https://github.com/MicroBuddha/Erythroblast.git}{\texttt{https://github.com/MicroBuddha/Erythroblast.git}}.

%
%
\bibliographystyle{splncs04}
\bibliography{final_manuscript}

\begin{thebibliography}{10}
\providecommand{\url}[1]{\texttt{#1}}
\providecommand{\urlprefix}{URL }
\providecommand{\doi}[1]{https://doi.org/#1}

\bibitem{abbafhs_classification_2018}
Abbas, K., Banks, J., Chandran, V., Tomeo-Reyes, I., Nguyen, K.: Classification of {White} {Blood} {Cell} {Types} from {Microscope} {Images}:{Techniques} and {Challenges}. pp. 17--25 (Nov 2018)

\bibitem{acevedo_dataset_2020}
Acevedo, A., Merino, A., Alférez, S., Ángel Molina, Boldú, L., Rodellar, J.: A dataset of microscopic peripheral blood cell images for development of automatic recognition systems. Data in Brief  \textbf{30},  105474 (2020). \doi{https://doi.org/10.1016/j.dib.2020.105474}, \url{https://www.sciencedirect.com/science/article/pii/S2352340920303681}

\bibitem{https://doi.org/10.1002/ajh.2830380209}
Adams, C.D., Kessler, J.F.: Circulating nucleated red blood cells following splenectomy in a patient with congenital dyserythropoietic anemia. American Journal of Hematology  \textbf{38}(2),  120--123 (1991). \doi{https://doi.org/10.1002/ajh.2830380209}, \url{https://onlinelibrary.wiley.com/doi/abs/10.1002/ajh.2830380209}

\bibitem{alkafrawi_blood_2022}
Alkafrawi, I.M.I., Dakhell, Z.A.: Blood {Cells} {Classification} {Using} {Deep} {Learning} {Technique}. In: 2022 {International} {Conference} on {Engineering} \& {MIS} ({ICEMIS}). pp.~1--6 (Jul 2022). \doi{10.1109/ICEMIS56295.2022.9914281}, \url{https://ieeexplore.ieee.org/document/9914281}

\bibitem{XGboost}
Chen, T., Guestrin, C.: Xgboost: A scalable tree boosting system. In: Proceedings of the 22nd ACM SIGKDD International Conference on Knowledge Discovery and Data Mining. KDD ’16, ACM (Aug 2016). \doi{10.1145/2939672.2939785}, \url{http://dx.doi.org/10.1145/2939672.2939785}

\bibitem{chola_bcnet_2022}
Chola, C., Muaad, A.Y., Bin~Heyat, M.B., Benifa, J.V.B., Naji, W.R., Hemachandran, K., Mahmoud, N.F., Samee, N.A., Al-Antari, M.A., Kadah, Y.M., Kim, T.S.: {BCNet}: {A} {Deep} {Learning} {Computer}-{Aided} {Diagnosis} {Framework} for {Human} {Peripheral} {Blood} {Cell} {Identification}. Diagnostics  \textbf{12}(11), ~2815 (Nov 2022), \url{https://www.mdpi.com/2075-4418/12/11/2815}, number: 11 Publisher: Multidisciplinary Digital Publishing Institute

\bibitem{KNN}
Cunningham, P., Delany, S.J.: k-nearest neighbour classifiers - a tutorial. ACM Computing Surveys  \textbf{54}(6),  1–25 (Jul 2021). \doi{10.1145/3459665}, \url{http://dx.doi.org/10.1145/3459665}

\bibitem{das_hematological_2016}
Das, R., Ahluwalia, J., Sachdeva, M.U.S.: Hematological {Practice} in {India}. Hematology/Oncology Clinics of North America  \textbf{30}(2),  433--444 (Apr 2016). \doi{10.1016/j.hoc.2015.11.009}, \url{https://www.sciencedirect.com/science/article/pii/S0889858815001963}

\bibitem{imagenet}
Deng, J., Dong, W., Socher, R., Li, L.J., Li, K., Fei-Fei, L.: Imagenet: A large-scale hierarchical image database. In: 2009 IEEE Conference on Computer Vision and Pattern Recognition. pp. 248--255 (2009). \doi{10.1109/CVPR.2009.5206848}

\bibitem{dosovitskiy2021imageworth16x16words}
Dosovitskiy, A., Beyer, L., Kolesnikov, A., Weissenborn, D., Zhai, X., Unterthiner, T., Dehghani, M., Minderer, M., Heigold, G., Gelly, S., Uszkoreit, J., Houlsby, N.: An image is worth 16x16 words: Transformers for image recognition at scale (2021), \url{https://arxiv.org/abs/2010.11929}

\bibitem{fang_fast_2022}
Fang, T., Yuan, P., Gong, C., Jiang, Y., Yu, Y., Shang, W., Tian, C., Ye, A.: Fast label-free recognition of {NRBCs} by deep-learning visual object detection and single-cell {Raman} spectroscopy. Analyst  \textbf{147}(9),  1961--1967 (May 2022). \doi{10.1039/D2AN00024E}, \url{https://pubs.rsc.org/en/content/articlelanding/2022/an/d2an00024e}, publisher: The Royal Society of Chemistry

\bibitem{he_deep_2015}
He, K., Zhang, X., Ren, S., Sun, J.: Deep {Residual} {Learning} for {Image} {Recognition} (Dec 2015). \doi{10.48550/arXiv.1512.03385}, \url{http://arxiv.org/abs/1512.03385}, arXiv:1512.03385 [cs]

\bibitem{SVM}
Hearst, M., Dumais, S., Osuna, E., Platt, J., Scholkopf, B.: Support vector machines. IEEE Intelligent Systems and their Applications  \textbf{13}(4),  18--28 (1998). \doi{10.1109/5254.708428}

\bibitem{howard2018universallanguagemodelfinetuning}
Howard, J., Ruder, S.: Universal language model fine-tuning for text classification (2018), \url{https://arxiv.org/abs/1801.06146}

\bibitem{huang2018denselyconnectedconvolutionalnetworks}
Huang, G., Liu, Z., van~der Maaten, L., Weinberger, K.Q.: Densely connected convolutional networks (2018), \url{https://arxiv.org/abs/1608.06993}

\bibitem{6115881}
Labati, R.D., Piuri, V., Scotti, F.: All-idb: The acute lymphoblastic leukemia image database for image processing. In: 2011 18th IEEE International Conference on Image Processing. pp. 2045--2048 (2011). \doi{10.1109/ICIP.2011.6115881}

\bibitem{RandomForest}
Louppe, G.: Understanding random forests: From theory to practice (2015)

\bibitem{nozaka}
NOZAKA, H., KUSHIBIKI, M., KAMATA, K., YAMAGATA, K.: Approach to recognition of immature granulocytes using deep learning in peripheral blood smear screening: The potential of ai models using a convolution neural network for blood cell morphology classification. Japanese Journal of Medical Technology  \textbf{73}(1),  69--77 (2024). \doi{10.14932/jamt.23-72}

\bibitem{paszke2019pytorch}
Paszke, A., Gross, S., Massa, F., Lerer, A., Bradbury, J., Chanan, G., Killeen, T., Lin, Z., Gimelshein, N., Antiga, L., Desmaison, A., Köpf, A., Yang, E., DeVito, Z., Raison, M., Tejani, A., Chilamkurthy, S., Steiner, B., Fang, L., Bai, J., Chintala, S.: Pytorch: An imperative style, high-performance deep learning library (2019)

\bibitem{pedregosa2018scikitlearn}
Pedregosa, F., Varoquaux, G., Gramfort, A., Michel, V., Thirion, B., Grisel, O., Blondel, M., Müller, A., Nothman, J., Louppe, G., Prettenhofer, P., Weiss, R., Dubourg, V., Vanderplas, J., Passos, A., Cournapeau, D., Brucher, M., Perrot, M., Édouard Duchesnay: Scikit-learn: Machine learning in python (2018)

\bibitem{rezatofighi_automatic_2011}
Rezatofighi, S.H., Soltanian-Zadeh, H.: Automatic recognition of five types of white blood cells in peripheral blood. Computerized Medical Imaging and Graphics: The Official Journal of the Computerized Medical Imaging Society  \textbf{35}(4),  333--343 (Jun 2011). \doi{10.1016/j.compmedimag.2011.01.003}

\bibitem{shekar}
Shekar, B.H., Dagnew, G.: Grid search-based hyperparameter tuning and classification of microarray cancer data. In: 2019 Second International Conference on Advanced Computational and Communication Paradigms (ICACCP). pp.~1--8 (2019). \doi{10.1109/ICACCP.2019.8882943}

\bibitem{simonyan_very_2015}
Simonyan, K., Zisserman, A.: Very {Deep} {Convolutional} {Networks} for {Large}-{Scale} {Image} {Recognition} (Apr 2015), \url{http://arxiv.org/abs/1409.1556}, arXiv:1409.1556 [cs]

\bibitem{sivarao_efficientnet_2023}
SivaRao, B.S.S., Rao, B.S.: {EfficientNet} - {XGBoost}: {An} {Effective} {White}-{Blood}-{Cell} {Segmentation} and {Classification} {Framework}. Nano Biomedicine and Engineering  \textbf{15}(2),  126--135 (Jun 2023). \doi{10.26599/NBE.2023.9290014}, \url{https://www.sciopen.com/article/10.26599/NBE.2023.9290014}

\bibitem{szegedy_going_2014}
Szegedy, C., Liu, W., Jia, Y., Sermanet, P., Reed, S., Anguelov, D., Erhan, D., Vanhoucke, V., Rabinovich, A.: Going {Deeper} with {Convolutions} (Sep 2014). \doi{10.48550/arXiv.1409.4842}, \url{http://arxiv.org/abs/1409.4842}, arXiv:1409.4842 [cs]

\end{thebibliography}

%




\end{document}